\documentclass[amsmath,prb,showpacs,twocolumn,superscriptaddress]{revtex4}

\usepackage{graphicx}

\begin{document}

\title{Molecular transport junctions: Current from
       electronic excitations in the leads}
\date{\today}
\author{Michael Galperin}
\affiliation{Department of Chemistry and Nanotechnology Center,
   Northwestern University, Evanston IL 60208}
\author{Abraham Nitzan}
\affiliation{School of Chemistry, The Sackler Faculty of Science,
   Tel Aviv University, Tel Aviv 69978, Israel}
\author{Mark A. Ratner}
\affiliation{Department of Chemistry and Nanotechnology Center,
   Northwestern University, Evanston IL 60208}

\begin{abstract}
Using a model comprising a 2-level bridge connecting free electron reservoirs 
we show that coupling of a molecular bridge to electron-hole excitations in the
leads can markedly effect the source-drain current through a molecular junction.
In some cases, e.g. molecules that exhibit strong charge transfer transitions, 
the contribution from electron-hole excitations can exceed the Landauer elastic 
current and dominate the observed conduction. 
\end{abstract}


\maketitle

%\section
\noindent
\textit{Introduction}.
Electron transport in molecular tunnel junctions has been the focus of intense 
recent research~\cite{ECIBooks,ANRev,HeathRatner,ReedBook}. Theoretical modeling
of tunnel conduction~\cite{ANMRRev,DattaBook} starts from Hamiltonians that 
contain electron transfer (tunneling) interactions between molecule and leads as essential 
elements for current transport in such junctions. At the same time, energy 
transfer interactions --- excitation/de-excitation of the molecule accompanied 
by electron-hole (EH) pair annihilation/creation in the metal --- are known to 
strongly affect the lifetime of excited molecules near metal 
surfaces~\cite{Silbey}.
An essential difference between these interactions is that electron transfer is
a tunneling process that depends exponentially on the molecule-metal distance, 
while energy transfer is associated with dipolar coupling that scales like the 
inverse cube of this distance, and can therefore dominate at larger distances.  

How will such dipolar interactions affect the conduction properties of 
molecular junctions? Here we address this question by using the non-equilibrium 
Green function (NEGF) formalism to derive an expression for the conduction in 
junction model that contains both electron and energy transfer interactions, 
then analyze several examples with reasonable parameters. We conclude that 
current caused by electron-hole excitations in the leads may be significant, 
sometimes even dominant, in situations when strong asymmetry, of a particular 
type explained below, in the molecule-lead coupling is present. As a simple 
extreme example consider the case where the highest occupied molecular orbital 
(HOMO) is coupled only to one lead, while the lowest unoccupied molecular 
orbital (LUMO) is coupled only to the other. Such a junction cannot pass 
current (in the absence of electronic correlations)
if the dipolar interaction is absent. Realistic situations will not be 
that extreme, still whenever the HOMO-LUMO transition is of the charge 
transfer type, we expect some degree of such asymmetry. We have recently 
shown~\cite{MGAN} that such situations may give rise to light induced current 
under zero voltage. We show below that also the current-voltage characteristic 
of such junctions is strongly affected by dipolar energy transfer 
interactions between molecule and leads.

%\section
\noindent
\textit{Model and Method}.
We consider a tunneling junction consisting of a molecule positioned between 
two metal contacts ($L$ and $R$). The molecule is represented by its highest 
occupied molecular orbital (HOMO), $|1>$, and lowest unoccupied molecular 
orbital (LUMO), $|2>$, with energies $\varepsilon_1$ and $\varepsilon_2$ and 
gap $\varepsilon_{21}=\varepsilon_2-\varepsilon_1$. The contacts are assumed 
to be free electron reservoirs, each at its own equilibrium, characterized by
electronic chemical potentials $\mu_L$ and $\mu_R$, where the difference 
$\mu_L-\mu_R=e\Phi$ is the imposed voltage. The corresponding Hamiltonian is
\begin{align}
 \label{H}
 &\hat H = \hat H_0 + \hat V_M + \hat V_N \\
 \label{H0}
 &\hat H_0 = \sum_{m=1,2}\varepsilon_m\hat c_m^\dagger\hat c_m
           +  \sum_{k\in\{L,R\}}\varepsilon_k\hat c_k^\dagger\hat c_k \\
 \label{VM}
 &\hat V_M = \sum_{K=L,R}\sum_{m=1,2;k\in K}\left(
              V_{km}^{(MK)}\hat c_k^\dagger\hat c_m+\text{H.c.}\right) \\
 \label{VN}
 &\hat V_N = \sum_{K=L,R}\sum_{k\neq k'\in K}\left(
              V_{kk'}^{(NK)}\hat c_k^\dagger\hat c_{k'}\hat c_2^\dagger\hat c_1
              +\text{H.c.}\right)
\end{align}
where $\text{H.c.}$ denotes Hermitian conjugate. Here the operators $\hat c_m$ 
and $\hat c_m^\dagger$ are annihilation and creation operators of electrons 
in the bridge ($m=1,2$), while $\hat c_k$ and $\hat c_k^\dagger$ are 
annihilation and creation operators of electrons in the leads. 
The Hamiltonian $\hat H_0$ is a sum of terms that correspond to the isolated 
molecule (represented by its HOMO-LUMO levels in our model) and free contacts. 
$\hat V_M$ describes the electron transfer (tunneling) process between these 
subsystems. This is the term usually employed to treat current in the biased 
junction. $\hat V_N$ represents coupling of the molecular HOMO-LUMO transition 
to electron-hole excitations in the contacts and is often used in models of
energy transfer between the molecule and the contacts. 

In the Keldysh NEGF formalism~\cite{Keldysh} the steady-state current through 
the junction is given by~\cite{MeirWingreen}
\begin{equation}
 \label{Isd}
 I_{sd} = \pm\frac{e}{\hbar} \int\frac{dE}{2\pi}\text{Tr}
 \left[\mathbf{\Sigma_{MK}^{<}}(E)\mathbf{G^{>}}(E)
      -\mathbf{\Sigma_{MK}^{>}}(E)\mathbf{G^{<}}(E)\right]
\end{equation}
calculated at the left ($K=L$ with ``$+$'' sign) or right ($K=R$ with ``$-$''
sign) contact, where the direction from left to right chosen positive.

The lesser and greater Green functions, $G^{<,>}$, needed in (\ref{Isd})
can be obtained from the Keldysh equation
\begin{equation}
 \label{G<>}
 \mathbf{G^{<,>}}(E)=\mathbf{G^r}(E)\,\mathbf{\Sigma^{<,>}}(E)\,\mathbf{G^a}(E)
\end{equation}
where the retarded and advanced Green functions, $G^{r,a}$, are given by 
the Dyson equation
\begin{equation}
 \label{Gra}
 \mathbf{G^{r}}(E) = \left[E-\mathbf{H_0^m}-\mathbf{\Sigma^r}(E)\right]^{-1};
 \quad\mathbf{G^{a}}(E)=\left[\mathbf{G^{r}}(E)\right]^\dagger
\end{equation}
Here $\mathbf{H_0^m}$ is a matrix that corresponds to the molecular part 
(first term on the right) of the Hamiltonian (\ref{H0})
and $\mathbf{\Sigma^r}(E)$ is the retarded self-energy matrix due to both 
direct and dipolar coupling to the leads. Here and below the matrices are 
given in the basis $\{|1>,|2>\}$. 

The self-energies needed in Eqs.(\ref{Isd})-(\ref{Gra}) are obtained within 
the usual diagrammatic technique on the Keldysh contour. In the non-crossing
approximation (NCA)~\cite{Bickers} this leads to
\begin{equation}
 \label{Sigma}
 \mathbf{\Sigma} = \mathbf{\Sigma^{(ML)}} + \mathbf{\Sigma^{(MR)}}
                 + \mathbf{\Sigma^{(NL)}} + \mathbf{\Sigma^{(NR)}}
\end{equation}
On the Keldysh contour these self energies are~\cite{Mahan,Jauho} $2\times 2$ 
matrices in the bridge space
\begin{align}
 \label{KCSigmaMKb}
 &\Sigma^{(MK)}_{mm'}(\tau_1,\tau_2) =
        \sum_{k\in K}V_{mk}^{(MK)}g_k(\tau_1,\tau_2)V_{km'}^{(MK)}
 \\
 \label{KCSigmaNK}
 &\Sigma^{(NK)}_{mm'}(\tau_1,\tau_2) = \delta_{mm'}\sum_{k\neq k'\in K}
 \left|V_{kk'}^{(NK)}\right|^2
 \nonumber\\* &\qquad\times
 g_k(\tau_2,\tau_1)g_{k'}(\tau_1,\tau_2) G_{\bar m\bar m}(\tau_1,\tau_2)
\end{align}
Here and below we use $\Sigma^{(PK)}_{mm'}$ to denote $mm'$ matrix element
($m,m'=1,2$) of the self energy, $P=M$ or $N$ corresponds to the
physical process (electron or energy transfer to the metal, respectively)
and $K=L$ or $R$ denotes the left and right leads, respectively.
$g_k$ is the free electron Green function 
in state $k$, and $\bar m=2\delta_{m,1}+\delta_{m,2}$, i.e. 
$\bar m=1$ if $m=2$ and vice versa.

After projection onto the real time axis we get the retarded, advanced,
lesser, and greater components of these self-energies, which in steady state 
situations can be expressed in energy space.
In the wide-band approximation~(see e.g. Ref.~\onlinecite{WBA}) the 
self-energies associated with electron exchange between molecule and leads have
the familiar forms~\cite{Jauho} 
\begin{subequations}
\label{SigmaMK}
\begin{align}
 \label{SigmaMKra}
 &\Sigma^{(MK)\,r}_{mm'} = \left[\Sigma^{(MK)\,a}_{m'm}\right]^{*}
 =-i\delta_{mm'}\Gamma^{(MK)}_{m}/2 \\
 \label{SigmaMK<}
 &\Sigma^{(MK)\,<}_{mm'} = i\delta_{mm'}f_K(E)\Gamma^{(MK)}_{m} \\
 \label{SigmaMK>}
 &\Sigma^{(MK)\,>}_{mm'} = -i\delta_{mm'}[1-f_K(E)]\Gamma^{(MK)}_{m} \\
 \label{GammaMK}
 &\Gamma^{(MK)}_{m} = 2\pi\sum_{k\in K}\left|V^{(MK)}_{km}\right|^2
                   \delta(E-\varepsilon_k) \\
 \label{fK}
 &f_K(E) = \left[\exp\left\{(E-\mu_K)/k_BT\right\}+1\right]^{-1}
\end{align}
\end{subequations}
(we neglect level mixing due to coupling to the contacts)
where $\mu_K$ is the chemical potential of the leads and $K=L,R$ 
denotes the left and right electrode, respectively.

The Langreth projection rules~\cite{Langreth} give the lesser and greater 
projections of the self-energies due to electron-hole excitations, 
Eq.(\ref{KCSigmaNK}), in the (diagonal) form~\cite{MGAN}
\begin{subequations}
\label{SigmaNK}
\begin{align}
 \label{SigmaNK<}
 &\Sigma^{(NK)\,<}_{mm}(E) = \int\frac{d\omega}{2\pi}
  B^{(K)}(\omega,\mu_K)G_{\bar m\bar m}^{<}(E+\omega) \\
 \label{SigmaNK>}
 &\Sigma^{(NK)\,>}_{mm}(E) = \int\frac{d\omega}{2\pi}
  B^{(K)}(\omega,\mu_K)G_{\bar m\bar m}^{>}(E-\omega) \\
 \label{BNKa}
 &B^{(K)}(\omega,\mu_K) = 2\pi\int dE\,
 \sum_{k\neq k'\in K}\left|V_{kk'}^{(NK)}\right|^2
 \\ &\quad\times
 \delta(E-\varepsilon_k)\delta(E+\omega-\varepsilon_{k'})
 f_K(E)[1-f_K(E+\omega)]
 \nonumber\\
 \label{BNKb}
 &\quad\equiv 2\pi \left|V^{(NK)}\right|^2 \rho_K^{e-h}(\omega)
\end{align}
\end{subequations}
Here $\rho_K^{e-h}(\omega)$ is density of electron-hole excitations
in the lead $K$, $\rho_K^{e-h}(\omega) = \int dE\, C^{(K)}(E,\omega) 
f_K(E)[1-f_K(E+\omega)]$, with
$C^{(K)}(E,\omega) = \sum_{k\neq k'\in K}
\delta(E-\varepsilon_k)\delta(E+\omega-\varepsilon_{k'})
 \approx \rho(E)\rho(E+\omega)$, where $\rho(E)$ is the density of lead
electronic states.
In the spirit of the wide-band approximation one can assume $C^{(K)}$ 
to be constant. This leads to
$\rho_K^{e-h}(\omega) = \omega\, C^{(K)}/\left[1-e^{-\beta\omega}\right]$.
Below we will use this expression in (\ref{BNKb}) 
to get $B^{(K)}(\omega,\mu_K)$.
The retarded and advanced self-energies, $\Sigma^{(NK)\, r,a}$, are
difficult to calculate from the Langreth rules. For simplicity we assume,
in the spirit of the wide band approximation, that all diagonal 
components of $\mathbf{\Sigma^{(NK)\, r}}$ and
$\mathbf{\Sigma^{(NK)\, a}}(E)= 
\left[\mathbf{\Sigma^{(NK)\, r}}(E)\right]^\dagger$
are purely imaginary. Then~\cite{MGAN}
\begin{equation}
 \label{SigmaNKra}
 \mathbf{\Sigma^{(NK)\, r}}(E) =\frac{\left[\mathbf{\Sigma^{(NK)\,>}}(E)
-\mathbf{\Sigma^{(NK)\,<}}(E)\right]}{2}
 \equiv -\frac{i}{2}\mathbf{\Gamma^{(NK)}}
\end{equation}

Eqs.~(\ref{G<>}), (\ref{Gra}), (\ref{SigmaNK<}), (\ref{SigmaNK>}) and
(\ref{SigmaNKra}) have to be solved self-consistently until convergence
is achieved. We use the level populations, 
$n_m = -i \int\frac{dE}{2\pi} G^{<}_{mm}(E)$ ($m=1,2$),
as a test for convergence. Convergence is declared when the population values 
at subsequent iteration steps change no more than the predefined tolerance,
taken below $10^{-6}$. 

The additive structure of the self-energy, Eq.(\ref{Sigma}), makes it possible
to separate the lesser and greater Green functions, Eq.(\ref{G<>}), and 
consequently also the source-drain current, Eq.(\ref{Isd}), into contributions 
due to direct electron transfer to the leads and coupling to the electron-hole 
excitations. 
\begin{equation}
 I_{sd}=I_{sd}^L+I_{sd}^{e-h}
\end{equation}
$I_{sd}^L$, is the usual Landauer expression for elastic current
\begin{align}
 \label{IL}
 I_{sd}^{L} &= \frac{e}{\hbar}\int_{-\infty}^{+\infty}\frac{dE}{2\pi}
 \sum_{m=1,2}\Gamma^{(ML)}_{m}G^r_{mm}(E)\Gamma^{(MR)}_{m}G^a_{mm}(E)
 \nonumber \\ &\times
 \left[f_L(E)-f_R(E)\right]
\end{align}
while $I_{sd}^{e-h}$ is the contribution from the electron-hole excitation. 
A simple expression for this current can be obtained when 
$\Gamma^{(MK)}_{m}\ll\varepsilon_{21}$, where $\Sigma^{(NK)\, <,>}$ takes an 
approximate form
\begin{subequations}
\label{appSigmaNK}
\begin{align}
 \label{appSigmaNK<}
 &\mathbf{\Sigma^{(NK)\,<}}(E) = i B^{(K)}
 \left[\begin{array}{cc}
 n_2 & 0 \\ 0 & 0
 \end{array}\right] 
 \\
 \label{appSigmaNK>}
 &\mathbf{\Sigma^{(NK)\,>}}(E) = -i B^{(K)}
 \left[\begin{array}{cc}
 0 & 0 \\ 0 & 1-n_1
 \end{array}\right]
\end{align}
\end{subequations}
Here $B^{(K)}$ is assumed to be a constant. Note that using $B^{(K)}$ as a 
parameter in the full self-consistent calculations below means that in fact we 
take $|V^{(NK)}|^2 C^{(K)}=B^{(K)}/2\pi\varepsilon_{21}$. 
Using (\ref{appSigmaNK}) in (\ref{Isd}) leads to the electron-hole excitations 
part of the flux in the form
\begin{align}
 \label{appIeh}
 &I_{sd}^{e-h} = \frac{e}{\hbar}B \times
  \\
  &\left[
  n_2^{(ML)}\left(\frac{\Gamma^{(MR)}_{1}}{\Gamma_1}-n_1^{(MR)}\right)
 -n_2^{(MR)}\left(\frac{\Gamma^{(ML)}_{1}}{\Gamma_1}-n_1^{(ML)}\right)\right]
 \nonumber
\end{align}
where $\Gamma_m= \Gamma^{(ML)}_{m} + \Gamma^{(MR)}_{m}
+ \Gamma^{(NL)}_{m} + \Gamma^{(NR)}_{m}$ 
($\Gamma_1^{(NK)}=B^{(K)}n_2$ and $\Gamma_2^{(NK)}=B^{(K)}[1-n_1]$)
and 
$n_m^{(MK)}= -i\int\frac{dE}{2\pi}\,\left|G_{mm}^r(E)\right|^2
\Sigma^{(MK)\, <}_{mm}(E)$ ($m=1,2$). 

\begin{figure}[htbp]
\includegraphics[width=\linewidth]{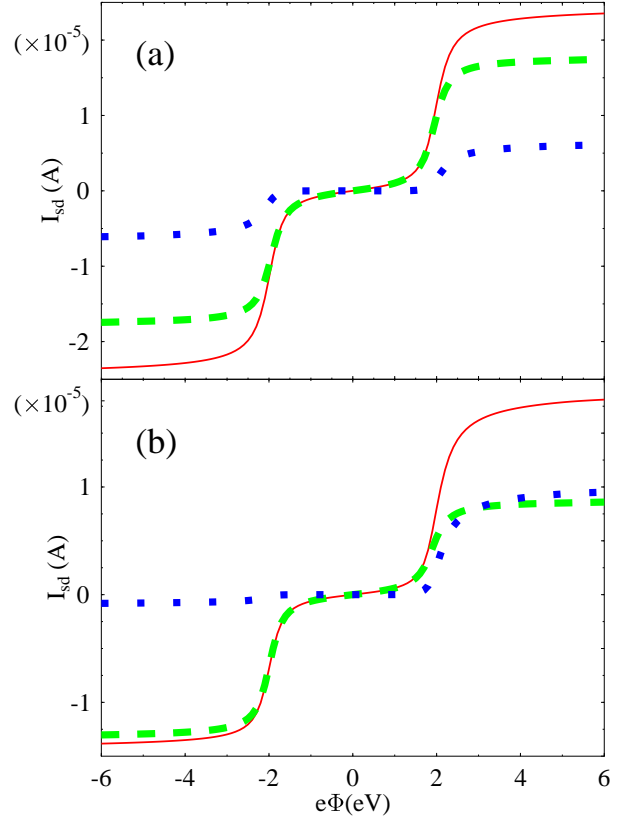}
\caption{\label{IPhi}The source-drain current $I_{sd}$ vs. applied voltage $\Phi$.
Shown are the total current $I_{sd}$ (full line, red) as well as contributions 
due to direct electron transfer $I_{sd}^{L}$ (dashed line, green) 
and electron-hole excitations $I_{sd}^{e-h}$ (dotted line, blue) for 
symmetric $\Gamma^{(ML)}_{1,2}=\Gamma^{(MR)}_{1,2}=0.1$~eV (a) 
and asymmetric $\Gamma^{(ML/R)}_{1}=0.1$~eV, $\Gamma^{(ML)}_{2}=0.19$~eV, 
and $\Gamma^{(MR)}_{2}=0.01$~eV (b) cases.}
\end{figure}

Further simplification of Eqs.~(\ref{IL}) and (\ref{appIeh})
is achieved for strong bias, e.g. for negatively biased left 
electrode where $\mu_L\gg\varepsilon_2$ and $\mu_R\ll\varepsilon_1$
so that $f_L=1$ and $f_R=0$ in the energy range relevant to the
integral in Eq.(\ref{IL}). Also in this case $n_m^{(MR)}=0$ and
$n_m^{(ML)}=\Gamma^{(ML)}_{m}/\Gamma_m$ ($m=1,2$) cam be used in (\ref{appIeh}).
Similar consideration apply in the opposite bias, leading finally to 
\begin{align}
 \label{simpleIL}
 &I_{sd}^{L} = \frac{e}{\hbar}
   \sum_{m=1,2}\frac{\Gamma^{(ML)}_{m}\Gamma^{(MR)}_{m}}{\Gamma_m} 
 \mbox{sgn}(\mu_L-\mu_R)\\
 \label{simpleIeh}
 &I_{sd}^{e-h} = \frac{e}{\hbar}B\times
 \\
 &\left[
 \frac{\Gamma^{(ML)}_{2}\Gamma^{(MR)}_{1}}{\Gamma_1\Gamma_2}\theta(\mu_L-\mu_R)
-\frac{\Gamma^{(ML)}_{1}\Gamma^{(MR)}_{2}}{\Gamma_1\Gamma_2}\theta(\mu_R-\mu_L)
 \right]
 \nonumber
\end{align} 
where $\theta$ is the step function $\theta(x)=1$ for $x>0$ and $0$ for $x<0$.
Note that $\left|I_{sd}^{e-h}\right|$ can be asymmetric to bias reversal
(see also Fig.~\ref{IPhi} and discussion below).
For, e.g., $\mu_L>\mu_R$
we see that the magnitude of $I_{sd}^{e-h}$ is determined both by the value 
of $B=B^{(L)}+B^{(R)}$ and by the product $\Gamma^{(ML)}_{2}\Gamma^{(MR)}_{1}$ 
while $I_{sd}^L$ is determined by the product 
$\Gamma^{(ML)}_{m}\Gamma^{(MR)}_{m}$ ($m=1,2$).
The electron-hole contribution to the source-drain current is significant 
when $\Gamma^{(ML)}_{2}>\Gamma^{(MR)}_{2}$ and/or 
$\Gamma^{(MR)}_{1}>\Gamma^{(ML)}_{1}$.
Below we compare the magnitude of the two contributions to the current for 
different junctions parameters.

%\section
\noindent
\textit{Numerical results}.
In the calculations reported below we used the following `standard' choice of 
parameters: $T=300$~K, $\varepsilon_{1}=0$~eV, $\varepsilon_{2}=2$~eV,
$\Gamma^{(M)}_{1}=\Gamma^{(M)}_{2}=0.2$~eV.
Values of other parameters are indicated in the figures.
The Fermi energy is taken in the middle of the HOMO-LUMO gap. 
Chemical potentials in the left and right leads 
are assumed to shift with the voltage bias symmetrically relative to the 
Fermi energy.
Numerical integration was done on the energy grid spanning range from $-3$ to 
$5$~eV with the step $10^{-3}$~eV. 

Figure~\ref{IPhi} depicts the current-voltage characteristic of the junction
for the cases of symmetric and asymmetric coupling between the molecular LUMO
and the contacts. Shown are the total current and its two components. 
In the symmetric case the current is dominated by the usual elastic electron 
(hole) transport through the LUMO (HOMO), and is symmetric relative to 
voltage reversal. 
The asymmetric case shows a significant contribution 
of the current associated with electron-hole excitations. 
The following points are noteworthy: (1) $I^{e-h}_{sd}$ is significant
when the LUMO is coupled asymmetrically to the two electrodes 
(2) This effect is particularly strong when the LUMO is coupled more
strongly to the negatively biased electrode (i.e. $\Phi<0$ when the LUMO 
couples strongly to the left). Indeed, $I_{sd}^{e-h}$ is expected to be 
pronounced when the LUMO is populated and the HOMO is empty, which 
happens at such bias.
Note also that the total current is asymmetric relative to bias reversal
in this case.

\begin{figure}[htbp]
\includegraphics[width=\linewidth]{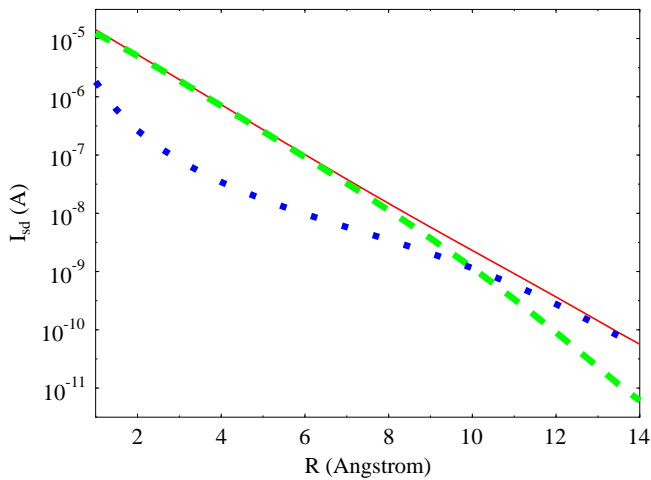}
\caption{\label{IR}Source-drain current $I_{sd}$ at $\Phi=3$~V vs.
molecule-contacts distance $R$. See text for choice of the coupling parameters.
Shown are the total $I_{sd}$ current (full line, red) as well as contributions 
due to direct electron transfer $I_{sd}^{L}$ (dashed line, green)
and electron-hole excitations $I_{sd}^{e-h}$ (dotted line, blue).}
\end{figure}

Figure~\ref{IR} shows the results of a model study of the dependence 
of the source-drain current on the molecule-lead distance $R$. 
We take $\Gamma^{(MK)}_{m} = A^{(MK)}_{m} \exp\left[-\alpha^{(MK)}_{m}R\right]$
to reflect a tunneling transition,
while $B^{(K)}(R)$ is assumed to have a dipolar distance 
dependence~\cite{Silbey},
$B^{(K)} = \beta^{(K)}/R^3$. 
The parameters used are $A^{(ML)}_{1}=A^{(MR)}_{1}=0.27$~eV, 
$A^{(ML)}_{2}=0.52$~eV, $A^{(MR)}_{2}=0.027$~eV, 
$\alpha^{(MK)}_{m}=1$~\AA${}^{-1}$, 
and $\beta^{(K)}=0.01$~eV~A${}^{3}$ ($K=L,R$ and $m=1,2$).
The choice of $A^{(MK)}_{m}$ reflects a total lifetime broadening for
electron transfer into the electrodes of $0.2$~eV at a distance 
(from each electrode) of $1$~\AA.
The choice of $\beta^{(K)}$ corresponds
to taking $B^{(K)}=0.01$~eV at this distance.
The relative importance of the $I_{sd}^{L}$ and $I_{sd}^{e-h}$ components
of $I_{sd}$ depends on the details of the molecule-leads
couplings. In particular, when $R\to\infty$ for both leads 
$\Gamma_m^{(M)}/B\to 0$ ($m=1,2$). In this limit $I_{sd}^{e-h}$ can become 
larger than $I_{sd}^{L}$.
As a specific example consider the situation where
$\Gamma_1^{(M)}=\Gamma_2^{(M)}$ and denote
$\xi_m=\Gamma^{(ML)}_m/\Gamma_m^{(M)}$. Then using Eqs.~(\ref{Sigma}), 
(\ref{appSigmaNK}), (\ref{simpleIL}), and (\ref{simpleIeh}) for the case
$\mu_R\ll\varepsilon_1<\varepsilon_2\ll\mu_L$
we obtain $I_{sd}^{e-h}/I_{sd}^{L}\to\xi_2/\xi_1(1-\xi_1-\xi_2)$,
$\infty$ and $(1-\xi_1)/(1-\xi_2)|1-\xi_1-\xi_2|$ when 
$\xi_2<1-\xi_1$, $\xi_2=1-\xi_1$ and $\xi_2>1-\xi_1$, respectively.
Our choice of parameters in Figure~\ref{IR} corresponds to the third case 
and yields ultimate dominance of $I_{sd}^{e-h}$
at large distances with $I^{e-h}_{sd}/I^L_{sd}\sim 100$ as $R\to\infty$.
Note that this limiting behavior is obtained only when both left and right 
molecule-metal couplings decrease together. 
Experimentally one of these distances can be controlled by moving a tip
while the other can be changed by adding insulating layers between molecule 
and substrate~\cite{Ho,Repp}.

%\section
\noindent
\textit{Conclusion}.
We have studied, within a simple model, the effect of dipolar energy-transfer 
interaction 
between molecule and leads on molecular conduction. We found that such 
interaction, that leads to electron-hole excitations in the contacts, can 
affect the current voltage characteristic of the junction in a substantial way 
and can not in general be disregarded. The contribution of this interaction can
dominate the overall conduction for particular asymmetric coupling where the 
molecular LUMO and/or HOMO are coupled differently to different leads. 
In addition, because of the different dependence of electron and energy 
transfer on the molecule-leads distance, the relative importance of
$I_{sd}^{L}$ and $I_{sd}^{e-h}$ depends on this distance, and can, 
in some cases, result in strong dominance of $I_{sd}^{e-h}$ at large 
molecule-lead separations.
 
\begin{acknowledgments}
We thank the NSF-NNI program, the DARPA Mol\-Apps initiative and the
Durint/MURI program of the DoD for support. A.N. thanks the Israel Science
Foundation for support.
\end{acknowledgments}

\end{document}